\documentclass{aa}
\usepackage{graphicx}

\def\***#1{***{\scshape #1}***}

\def\ergs{erg s$^{-1}$ cm$^{-2}$}

\def\apj{ApJ}
\def\nat{Nature}

\begin{document}

\sloppypar

   \title{The  broad-band spectrum of the persistent emission from SGR~1806-20}

   \author{S. Molkov$^{1,3}$, K. Hurley$^{2}$, R. Sunyaev $^{1,3}$,
   P. Shtykovsky $^{1,3}$, M. Revnivtsev $^{1,3}$}
 
   \offprints{molkov@hea.iki.rssi.ru}

   \institute{Space Research Institute, Russian Academy of
              Sciences, Profsouznaya 84/32, 117997 Moscow, Russia
        \and UC Berkeley Space Sciences Laboratory, 7 GAUSS WAY, Berkeley, CA 94720-7450, USA 
        \and Max-Plank-Institute fur Astrophysik,
              Karl-Schwarzschild-Str. 1, 85740 Garching-bei-Munchen, Germany
        }
  \date{}

  \authorrunning{Molkov et al.}
  \titlerunning{}
        
\abstract{We present the results of an analysis of the quiescent X-ray
emission  from the
Soft Gamma-Ray Repeater SGR~1806-20, taken during an INTEGRAL ultra-deep
survey of the
Galactic Center region in autumn 2003. The total effective exposure time 
spent on the source by the IBIS telescope during these observations 
exceeded 1.5 million seconds. Combining the INTEGRAL results with 
results from the XMM-Newton observatory, we present the broad
band (1-200 keV) spectrum of the quiescent emission from this source.
This is the first spectrum of the persistent emission
from an SGR in the broad energy range up to 200 keV. 
The luminosity of SGR~1806-20 in this range was
$\sim 3.6\times10^{36}$ ergs s$^{-1}$ for
an assumed distance of 15 kpc. We show that weak undetected bursts should
not contribute significantly to the quiescent emission.
The spectrum of the source is very hard and has a  power law
shape without any trace of a high energy cutoff up to $\sim$160 keV.
No strong cyclotron line was detected in the persistent
spectrum in the previously reported 4-6 keV band. During our next observations
in the autumn 2004 source went to active phase and its avereged
flux between powerful bursts was 2-3 times higher than in 2003.
During these observations two other SGR candidates, SGR~1801-23 and SGR~1808-20,
were in the field of view. Neither persistent hard X-ray emission nor
bursts were detected from them. The upper limit on the persistent flux
from each of them in the energy band
18-100 keV is about $4\times10^{-11}$ \ergs.
}

\maketitle

%

\section{Introduction}

Soft gamma repeaters (SGRs) are rare objects
which have been discovered by their sporadic bursting activity. The
bursts may last anywhere from $\sim$ 100 ms to many minutes in a few
cases, and their spectra may extend up to $\sim$ 100 keV or beyond.
These properties make them easily observable for
small, isotropic hard X-ray and soft gamma ray experiments
(see e.g. \cite{mazets81a}, Hurley 2000, 
\cite{aptekar2001}, \cite{woods2004} for reviews). 
In particular, SGR~1806-20 was discovered and localized
on January 7, 1979 during a single 
short (duration less than a quarter of second) flare in soft gamma rays
(\cite{mazets1981}, \cite{laros1986}). Numerous bursts have been
detected from this source since then (\cite{hurley1986},
\cite{laros1987}, \cite{kuznetsov1987}, 
\cite{atteia1987}, \cite{kouveliotou1987}) and intermittent bursting activity has
continued up to the present day. 

The energy emitted by SGRs is in many cases significantly larger than that 
available from the rotational spin-down. Therefore an additional energy 
reservoir is needed to power the observed emission. The most widely
accepted model for SGRs suggests that this power comes not from accretion 
but rather due to the decay of
an ultra-strong magnetic field of an isolated neutron star
(the ``magnetar'' model, Paczy\'nski 1992; Duncan \& Thompson 1992;
Thompson \& Duncan 1995). 

Studies of the four known SGRs
in soft X-rays (2-10 keV) have revealed that they emit a persistent
X-ray flux
(\cite{hurley1996}, \cite{hurley1999}, \cite{kulkarni2003},
\cite{kouveliotou2003}, \cite{mereghetti2000}, \cite{woods2001}).
The study of their persistent hard X-ray emission is more difficult, 
because of the weak fluxes and the locations of the sources in crowded fields.
The INTEGRAL mission, however, with its imaging detectors operating in
the hard X-ray and gamma-ray ranges (\cite{winkler03}), has made the
study of these sources
possible up to energies which have not been explored before.

In this letter we present the broad band spectrum (1-200 keV) of the persistent
emission of SGR~1806-20 obtained with long exposures by the INTEGRAL
observatory in autumn 2003. The source flux was
$\sim 1.3\times 10^{-10}$ erg s$^{-1}$ cm$^{-2}$,
which corresponds to a source luminosity 
$\sim 3.6\times10^{36}$ erg s$^{-1}$ in this energy band (assuming a source 
distance of 15 kpc). 
In a companion paper, Mereghetti et al. (2004) present independent evidence,
from a separate observation, of hard, persistent X-ray emission from this source.
Note that the persistent emission of SGR 1806-20
varies with time. For example, during recent 2004  observations
The RXTE/PCA (\cite{woods2004b}) and INTEGRAL (\cite{molkov2004a})
detected a high level of burst activity and 2-3 times more persistent
flux in the hard X-rays (\cite{molkov2004b}) than in the autumn of 2003.

\begin{figure}[t]
 \vspace{-1.8cm}
 \includegraphics[width=7.5cm]{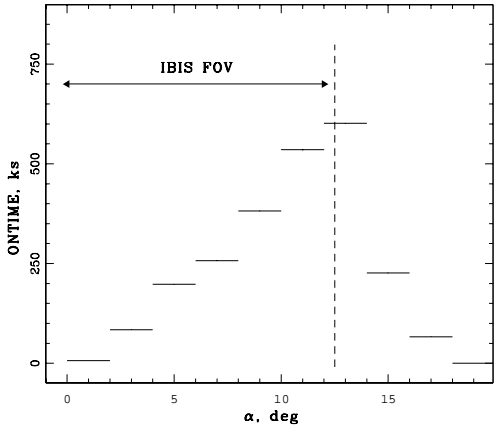}

 \vspace{-0.3cm}

 \caption{The differential distribution of IBIS exposure time on SGR~1806-20
   as a function of the angle between the optical axis of the instrument
   and the direction to the source. The dashed line denotes the half
angle
   of the IBIS Field of View (FOV). The total exposure time is
1.6 Msec.
} 
 \vspace{-0.2cm}
 \label {distr}
\end{figure}

\section{Observations and data analysis}

SGR 1806-20 is located $\sim10^{\circ}$ from the Galactic Center
and it was therefore observed by the wide field telescopes
aboard INTEGRAL (IBIS and SPI), but was outside the field of view
of JEM-X
during the ultra-deep survey of the Galactic Center field
performed during the Russian share of the INTEGRAL Open time AO-1
observing program (ID0120213; August 23 - September 24, 2003).
The total exposure time of these observations is $\sim2$Msec.
In order to  distinguish the flux of
SGR~1806-20 from that of nearby bright X-ray sources we used only data from 
the IBIS/ISGRI telescope (\cite{uber03}) which has moderate angular resolution
($\sim 12^\prime$). The SPI telescope has angular resolution $\sim 
2-2.5^\circ$ and cannot work in such a crowded field.
IBIS/ISGRI effectively operates in the energy range 18--400 keV
with typical sensitivity $\sim$1.5 mCrab for one orbit ($\sim3$
days) of observations. The fully coded field of view of this telescope is 
$10^\circ \times 10^\circ$, and the zero response field of view is 
$29^\circ \times 29^\circ$.
Due to the fact that the source was practically always  outside of the fully
coded field of view (the region of maximum effective area of the telescope)
the effective exposure is lower than the total exposure time, namely
$\sim 1.6$Msec.
The distribution of exposure time as a function of offset angle
is presented
in Fig.\ref{distr}

The IBIS/ISGRI data analysis was done with the software developed by Eugene
Churazov at the Space Research Institute, Moscow  and described in the
paper of
Revnivtsev et al. (2004). For the spectral reconstruction
we used  the ratio of the fluxes measured in  various energy
channels to the fluxes measured by the ISGRI detector from the Crab
Nebula in the same energy bands; wherever possible the ratios were calculated
for the same positions in the telescope field of view.

In order to extend the spectrum of SGR 1806-20 to lower energies
we have used the data from an XMM-Newton observation performed on
October 7, 2003,
i.e. practically simultaneously with our observation.
In the analysis we have used the data of
the PN and MOS1 cameras, which were processed with the Science
Analysis System (SAS) v6.0.0.

\section{Results}

\subsection{SGR 1806-20}

The first deep IBIS/ISGRI map of Galactic Center region was presented
in the work \cite{revnivtsev04}. SGR~1806-20 
was detected in the 18-60 keV energy range, along with sixty other sources.
It was the first report of the detection of this source in hard X-rays.

\begin{figure}[t]
 \vspace{-0.cm}
 \includegraphics[width=\columnwidth]{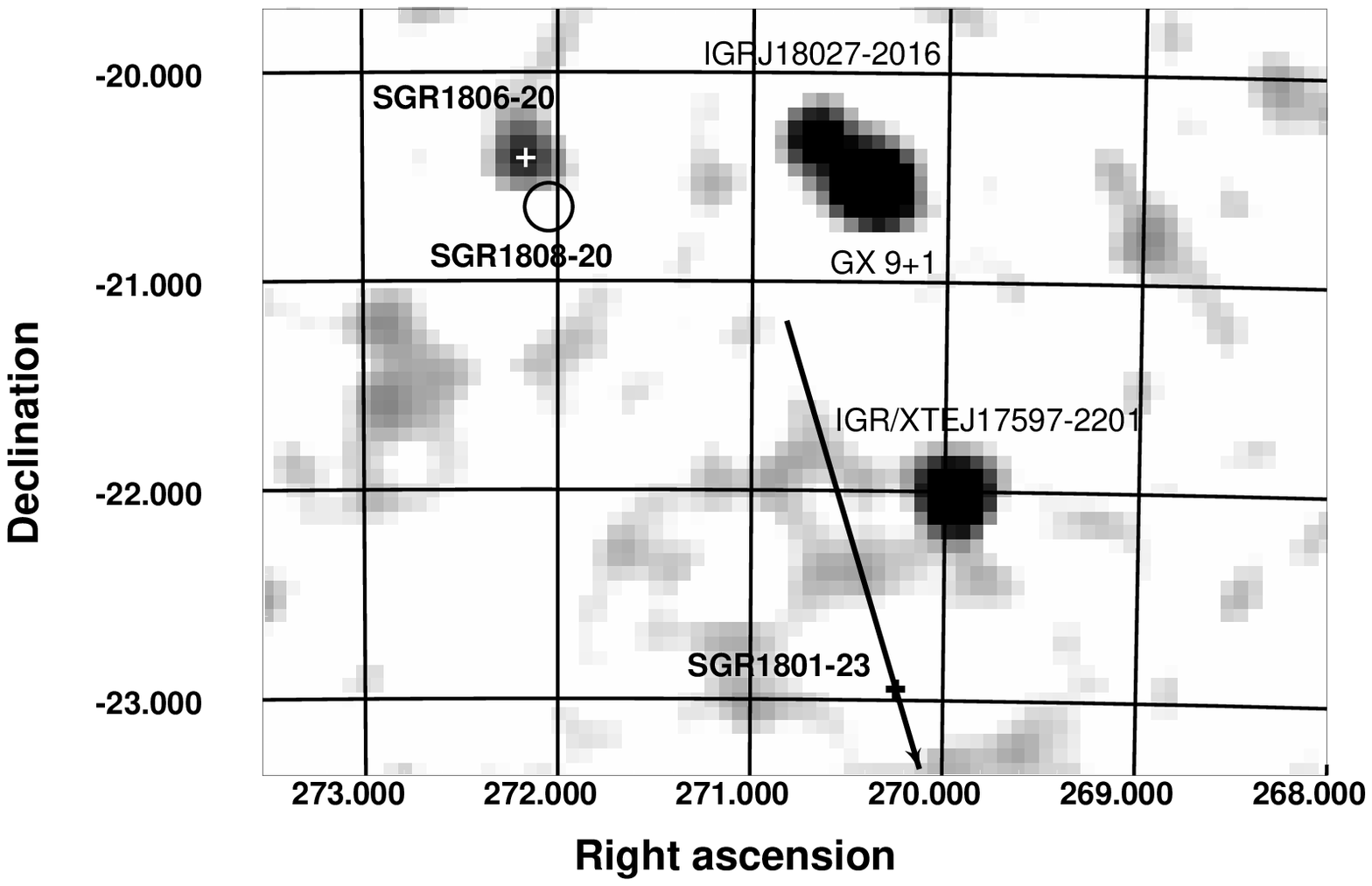}

 \vspace{-1.7cm}

 \includegraphics[width=\columnwidth]{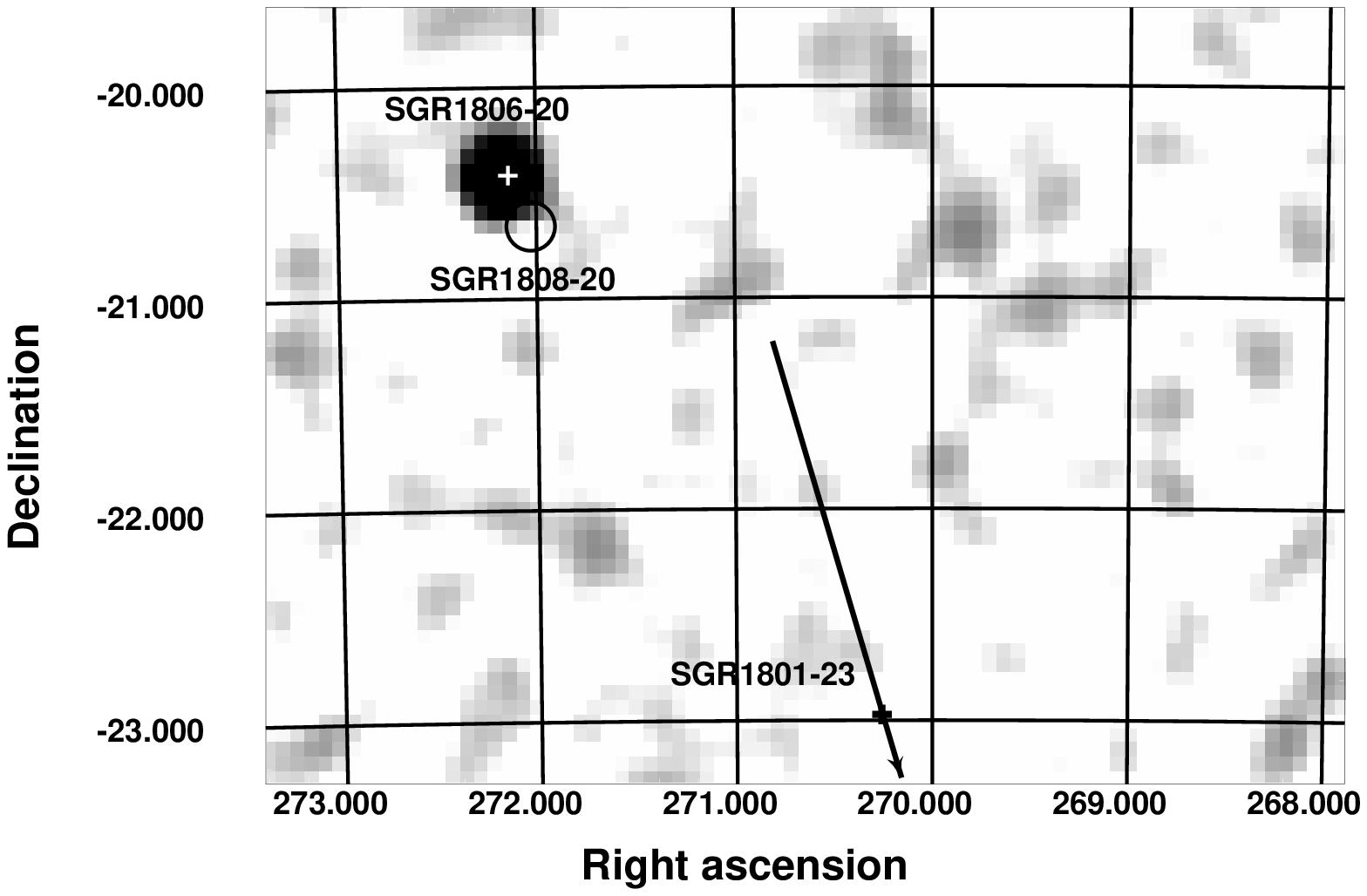}

 \vspace{-1.cm}

 \caption{Top panel: time averaged image of the region around SGR 1806-20
Bottom panel: image constructed during the bursts of SGR 1806-20
 detected in the ultra-deep autumn
   2003 observations of the Galactic center region.
 Both images are reconstructed in 18-60 keV energy band. {The
entire error boxes of the sources SGR~1808-20 and SGR~1801-23 were in
the field of view, but neither persistent
emission nor bursts from
these sources was detected during our observations.} } 
 \label {local}
\end{figure}

During our INTEGRAL observations the source 
burst numerous times (\cite{sunyaev2003}, G\"otz et al. 2004).
In order to demonstrate
that the origin of these bursts was SGR~1806-20 and not from other
possible candidates (SGR~1808-20, \cite{lamb2003} or SGR 1801-23, 
\cite{cline2000}) in the wide field of view of  IBIS/ISGRI
we present the  $\sim3^\circ\times2^\circ$ image around SGR~1806-20,
collected during all the bursts which were detected during our observations.
Detailed analysis of these bursts and others detected by INTEGRAL will
be presented
in a separate paper. On the lower panel of
Fig.\ref{local} we present the time-averaged map for the same region.
Note that SGR~1806-20 is clearly detected.

The burst and quiescent emission of SGR~806-20 are believed to originate
from different regions on and around the neutron star, and are thought
to be due to different mechanisms.  Therefore, to obtain the spectrum of
the quiescent emission, it is necessary to eliminate any possible contribution
from bursts.

\begin{figure}[t]

 \vspace{-0.5cm}

\includegraphics[width=\columnwidth]{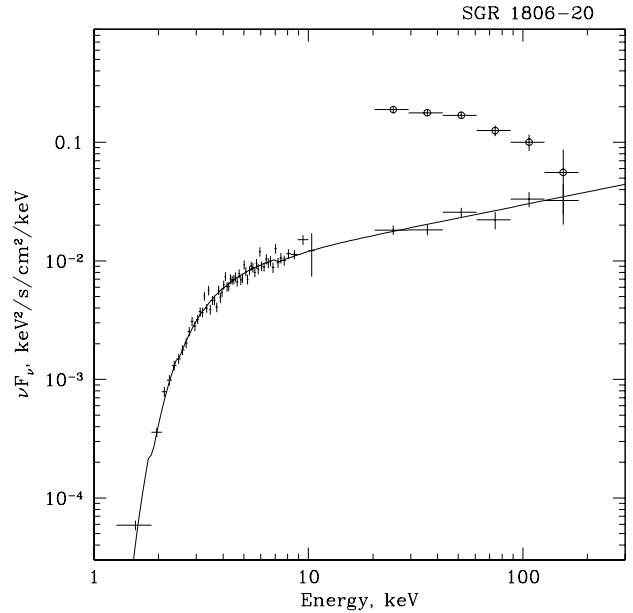}

 \vspace{-0.5cm}

 \caption{The quiescent energy spectrum of SGR 1806-20 (lower spectrum) 
and the summed spectrum of all detected bursts rescaled down by a factor 
of 1000 (upper spectrum)} 

 \vspace{-0.2cm}

 \label {spc}
\end{figure}

To do this, we constructed the statistical distribution of
the detector count rates with time resolution 0.01 s. We
identified candidate bursts as time intervals where the count rate
during at least four consecutive bins exceeded the mean value
of the count rate by more than four times the RMS value. Only those
candidate bursts which demonstrated
the clear presence of a point-like source on the image constructed during the
burst intervals were considered as real. All the ``real'' bursts detected
during our observations originated from 
SGR~1806-20. Their typical duration  was $50-200$ msec.
The  minimum number of counts from the source during
these bursts was  40-50, which corresponds to
a minimum fluence $\sim 2-3\times10^{-8}$ erg cm$^{-2}$. 
The majority of the remaining candidate events are  probably
caused by the anomalous short-term activation of one of the detector
pixels  (the so-called ``short-lived hot pixels'').
During these intervals more than 50\% of all detected counts come from
only one
pixel, while for ``real'' bursts this fraction is typically less than 10\%. 
Also, these bursts did not produce a point source image. 

The summed spectrum of all the ``real'' bursts is presented in
Fig.\ref{spc}. The best fit
spectrum using an optically
thin thermal plasma fit ($bremss$ model in XSPEC fitting package) has
$\rm kT \sim 47\pm5$, which is consistent with that presented
by G\"otz et al. (2003)
for one of the brightest bursts detected in a different set of observations.

We have filtered out the detected bursts from our data and have constructed 
the time-averaged spectrum of the total (i.e. pulsed plus unpulsed) persistent
emission from SGR 1806-20. 
In order to complement this hard X-ray 
spectrum with the spectrum at lower, X-ray energies
we have used the
publicly available data from the XMM-Newton (EPIC-pn and EPIC-MOS1)
observations of the source carried out on October 7, 2003. The broad-
band spectrum of SGR~1806-20
is presented in Fig. \ref{spc}. In this figure we have rescaled the
XMM spectrum
by 35\% in order to match the INTEGRAL/IBIS normalization.
This discrepancy in the normalization factor  could be partially caused by
the intrinsic source variability, but it is likely that most of the
difference originates in the uncertain cross-calibration of the INTEGRAL
and XMM instruments (see e.g. \cite{courvoisier03}).
The spectrum can be well described by a single power law with 
photoabsorption (presumably interstellar). We have not detected statistically 
significant temporal variations of the source spectral shape in our
observations. The best fit photon index for the time-averaged spectrum
between 18 and 200 keV 
is $\Gamma=1.6\pm0.1$ and the absorption column 
$N_{\rm H}=(6.6\pm0.2)\times 10^{22}$ cm$^{-2}$. A lower limit on a possible
exponential cutoff in the spectrum  of the form $dN/dE\propto E^{-\Gamma} 
\exp(-E/E_{\rm cut})$ is $E_{\rm cutoff}>160$ keV
(2$\sigma$ lower limit). The reduced $\chi2$ for 
the best fit model is $\chi2_r=$1.15 for
294 degrees of freedom. The total observed flux in the 1-200 keV
energy band is $\sim 1.3\times 10^{-10}$ erg s$^{-1}$ cm$^{-2}$ (not
corrected for absorption). This is approximately 8 times higher than 
the source emissivity in X-rays $<10$ keV.
For an assumed source distance of 15 kpc (\cite{corbel04}) the
observed
broad-band source flux corresponds to a 
luminosity $\sim 3.6 \times 10^{36}$ erg s$^{-1}$. The energy loss due
to the spin down of the neutron star in SGR~1806-20 is approximately
three orders of magnitude smaller, namely 
$\dot{E}_{\rm rot}\sim 10^{33}$ erg s$^{-1}$ (\cite{kouveliotou1998}).
Obviously the rotational energy loss cannot power either the
observed X-ray emission or the  hard X-ray (20-200 keV) emission.

The RXTE/PCA  has detected a relatively strong narrow
line at 5 keV during some bright bursts, interpreted as cyclotron absorbtion 
 (\cite{ibrahim02,ibrahim03}).
The high quality 1-10 keV XMM spectrum of SGR~1806-20
allows us to put constraints on the possible presence of 
such a feature in the persistent spectrum of
the source during this observation.
Fixing the energy of the line to the 
4.0-6.0 keV range, we can place an upper limit on a possible absorption line
component in the spectrum. The 2$\sigma$ upper limit on the line equivalent 
width varies $EW\la 50$ eV for 0.1 keV width of the line and $EW\la$160 eV
for 0.5 keV width of the line. This is
significantly smaller than the equivalent width of the line detected
by the RXTE/PCA during the bursts, $EW\sim 500-800$ eV.
The large difference
between the two equivalent widths may be connected with differences
in the physical conditions
leading to formation of burst and persistent spectra.

\subsection{SGR~1801-23 and SGR~1808-20}

During our INTEGRAL observations two SGR candidates 
were in the field of view. Both
sources have rather large  location uncertainties.
SGR~1808-20 is localized to  a $\sim6$ arcmin radius circle and
SGR~1801-23 to a $\sim200\times0.5$ arcmin error box.
The error boxes of the both sources were entirely in the IBIS field of
view (see the Fig.\ref{local}).
Neither persistent nor bursting emission was detected
from these sources.  The upper limit on the persistent flux from
these sources is about $4\times10^{-11}$ \ergs. 

\section{Discussion}

Although we do not know the exact mechanism which produces the
persistent emission of SGR~1806-20, we can nevertheless exclude
a significant contribution to it from weak, undetected bursts by
the following argument.
We can estimate the contribution using the 
number-intensity distribution
of the bursts  
presented in  Gogus et al. (2000) and G\"otz et al. 
(2004). The peak number of detected bursts with fluence $\ga 2-3
\times 10^{-8}$ ergs cm$^{-2}$ is approximately 20-30 bursts per 10 days. 
The total flux that bursts with fluences lower than this
can produce during the exposure time of our observations
is not more than 
a few$\times 10^{-13}$ erg s$^{-1}$ cm$^{-2}$ if we assume that 
the number-fluence distribution of bursts from SGR 1806-20 
can be described by a function $N(>F)\propto F^{-0.7}$ (see \cite{gogus00}, 
\cite{goetz04}). This flux does not contribute even one percent to
the detected persistent flux of  SGR~1806-20. Therefore
we can conclude that the detected quiescent spectrum of
SGR~1806-20 cannot be strongly affected by weak undetected bursts.
Additional indirect support for this conclusion comes from 
the completely different spectral shapes of the bright bursts
and the quiescent emission detected during our observations
(see Fig.\ref{spc}).

\bigskip

{\it Acknowledgments.} The authors thank E.Churazov for the development of
the analysis methods for the IBIS data and software.
This research
has made use of data obtained through the INTEGRAL Science Data Center
(ISDC), Versoix, Switzerland, Russian INTEGRAL Science Data Center (RSDC),
Moscow, Russia. KH is grateful for support under the INTEGRAL U.S. guest
investigator program, NASA grant NAG5-13738.

\end{document}